\documentclass[twocolumn,preprintnumbers,elsart]{revtex4}
\usepackage{makeidx}
\usepackage{amssymb}
\usepackage{amsmath}
\usepackage{mathrsfs}
\usepackage{graphicx}
\usepackage{dcolumn}
\usepackage{bm}
\usepackage[center]{subfigure}
\usepackage{color}
\usepackage{indentfirst}
\usepackage{array}
\usepackage{booktabs}
\usepackage{multirow}

\begin{document}

\title{Vortex solitons in quasi-phase-matched photonic crystals}
\author{Feiyan Zhao$^{1}$}
\author{Xiaoxi Xu$^{1}$}
\author{Hexiang He$^{1}$}
\author{Li Zhang$^{1}$}
\author{Yangui Zhou$^{1}$}
\author{Zhaopin Chen$^{2}$}
\author{Boris A. Malomed$^{3,4}$}
\author{Yongyao Li$^{1,5}$}
\email{yongyaoli@gmail.com}
\affiliation{$^{1}$School of Physics and Optoelectronic Engineering, Foshan University,
Foshan 528000, China\\
$^{2}$Physics Department and Solid-State Institute, Technion, Haifa 32000,
Israel\\
$^{3}$Department of Physical Electronics, School of Electrical Engineering,
Faculty of Engineering, Tel Aviv University, Tel Aviv 69978, Israel \\
$^{4}$Instituto de Alta Investigaci\'{o}n, Universidad de Tarapac\'{a},
Casilla 7D, Arica, Chile\\
$^{5}$Guangdong-Hong Kong-Macao Joint Laboratory for Intelligent Micro-Nano
Optoelectronic Technology, Foshan University, Foshan 528000, China}

\begin{abstract}
We report solutions for stable compound solitons in a three-dimensional
quasi-phase-matched photonic crystal with the quadratic ($\chi ^{(2)}$)
nonlinearity. The photonic crystal is introduced with a checkerboard
structure, which can be realized by means of the available technology. The
solitons are built as four-peak vortex modes of two types, rhombuses and
squares (intersite- and onsite-centered self-trapped states, respectively).
Their stability areas are identified in the system's parametric space
(rhombuses occupy an essentially broader stability domain), while all bright
vortex solitons are subject to strong azimuthal instability in uniform $\chi
^{(2)}$ media. Possibilities for experimental realization of the solitons
are outlined.
\end{abstract}

\author{}
\maketitle


Quadratic ($\chi ^{(2)}$) media provide a versatile platform for the
creation of optical solitons \cite{Lederer,1Buryak2002}. Unlike the
self-focusing cubic ($\chi ^{(3)}$) nonlinearity, which generates
two-dimensional (2D) solitons that are subject to the instability driven by
the critical collapse in this setting \cite{2Fibich2002,book}, $\chi ^{(2)}$
systems support stable 2D fundamental (zero-vorticity) solitons in free
space \cite{1Buryak2002,3Torruellas1995,4Torruellas2002}. Actually, the $%
\chi ^{(3)}$ effect can be emulated in $\chi ^{(2)}$ media by means of the
cascading mechanism when the intensity of the light beam increases to $\sim
10$ GW/cm$^{2}$ \cite{5DeSalvo1992,6Bosshard1995}.

Unlike the fundamental 2D $\chi ^{(2)}$ solitons, ones with embedded
vorticity (alias vortex rings) exhibit strong azimuthal instability which
splits the vortex rings into fragments \cite{Firth,Torner1,Torner2}, which
makes stabilization of vortex solitons a fundamental problem \cite{book}.
Working in this direction, it was predicted \cite{7Towers2001,10Mihalache}
and demonstrated experimentally \cite{8Trapani2000} that families of stable
2D vortex solitons can be supported by the competition of $\chi ^{(2)}$ and
self-defocusing $\chi ^{(3)}$ nonlinearities. In other settings, stability
of solitary vortices is provided by competing nonlinearities of other types,
such as cubic-quintic in various forms \cite%
{Michinel,Skarka,Mihalache2002,Pego,Mihalache2006,Soto-Crespo,Soto-Crespo2,Brand,NonlinDyn}%
. cubic-quartic in 3D \cite{YVK2018}, and cubic times a logarithmic factor
in 2D \cite{YLi2018}. In the latter two cases, such combinations of
nonlinear terms model the competition of mean-field and beyond-mean-field
interactions in binary Bose-Einstein condensates \cite{Petrov1,Petrov2}.
Recently, stable vortex dissipative solitons have been predicted in a pure $%
\chi ^{(2)}$ medium, by adding ring-shaped gain to the spatially uniform
lossy background \cite{11Lobanov2022}.

\begin{figure}[h]
{\includegraphics[width=0.8\columnwidth]{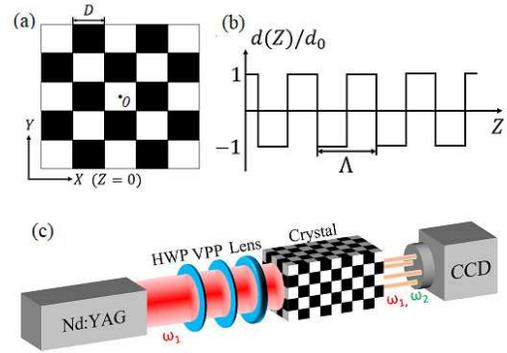}}
\caption{(a) The checkerboard structure representing the nonlinear lattice
in the input plane ($Z=0$) as per Eq. (\protect\ref{checkerboard}). Black and white color cells correspond to $\sigma(X,Y)=1$ and $-1$, respectively. (b) The
modulation of $d(Z)$, at a fixed point, where $\protect\sigma (X,Y)=+1$ (black color cells),
along the propagation distance with period $\Lambda $. (c) A schematic of
the proposed experimental setup.}
\label{crystal}
\end{figure}

Designing new setups admitting stability of vortex solitons in $\chi ^{(2)}$
media remains a relevant problem -- in particular, because this may open a
way for the creation of robust vortex rings using low-power optical beams.
In this work, we propose a new possibility: stabilization of vortex rings by
means of quasi-phase matching (QPM) nonlinear photonic crystals. By itself,
QPM is a well-known method for achieving accurate phase matching in $\chi
^{(2)}$ crystals for the nonlinear frequency conversion \cite{9Arie} in
various settings \cite{HLi2020}. This technique is based on spatial
modulation of the $\chi ^{(2)}$ susceptibility, imposed by means of periodic
poling of the crystalline material \cite{poling1,poling2,poling3} in 1D/2D
and the femtosecond laser engineering technology in 3D \cite%
{Tianxiang2019,Dwei2018,Tal2019,Ady2021}, while the linear susceptibility is
kept constant. Generally, there are three parameters of the QPM structure:
the modulation period, duty cycle, and phase shift \cite{Aviv2018}. A
periodically distributed phase shift creates a nonlinear lattice, alias a
nonlinear photonic crystal \cite{JYang2009}.

We here elaborate a mechanism for the stabilization of vortex solitons by a 3D QPM nonlinear photonic crystal. The paraxial propagation of light along direction $Z$ is
governed by coupled equations for the slowly varying fundamental-frequency
(FF) and second-harmonic (SH) amplitudes, $A_{1}$ and $A_{2}$:
\begin{align}
& i\partial _{Z}A_{1}=-\frac{1}{2k_{1}}\nabla ^{2}A_{1}-\frac{%
2d(X,Y,Z)\omega _{1}}{cn_{1}}e^{-i\Delta kZ}A_{1}^{\ast }A_{2},  \label{A1}
\\
& i\partial _{Z}A_{2}=-\frac{1}{2k_{2}}\nabla ^{2}A_{2}-\frac{d(X,Y,Z)\omega
_{2}}{cn_{2}}e^{i\Delta kZ}A_{1}^{2},  \label{A2}
\end{align}%
where $\nabla ^{2}=\partial _{X}^{2}+\partial _{Y}^{2}$ is the
paraxial-diffraction operator, $c$ is the speed of light in vacuum, $k_{1,2}$%
, $\omega _{1,2}$ ($\omega _{2}=2\omega _{1}$), and $n_{1,2}$ are,
respectively, the FF and SH\ carrier wavenumbers, frequencies, and
refractive indices, with the phase mismatch $\Delta k=2k_{1}-k_{2}$.
Further, the 3D periodically modulated local
$\chi ^{(2)}$ susceptibility is determined by $d(X,Y,Z)=\sigma (X,Y)d(Z)$, with a checkerboard structure in the $(X,Y)$ plane:
\begin{equation}
\sigma (X,Y)=-\mathrm{sgn}\left\{ {\cos (\pi X/D)\cos (\pi Y/D)}\right\} ,
\label{checkerboard}
\end{equation}%
where $D\times D$ is the size of each square cell, as shown in Fig. \ref{crystal}(a) (a 1D version of this pattern was
considered to spatial control of entangled two-photon states, in Ref. \cite{Torres2004}), and $d(Z)=d_{0}\mathrm{sgn}[\cos(2\pi Z/\Lambda)]$ is modulation function along the propagation distance, as shown in Fig. \ref{crystal}(b), which can be represented by its Fourier expansion corresponding to modulation period $\Lambda$:
\begin{equation}
d(Z)=d_{0}\sum_{m\neq 0}\frac{2}{\pi m}{\sin }\left( \frac{\pi m}{2}\right) {%
\exp \left( i\frac{2\pi mZ}{\Lambda }\right) ,}  \label{QPM}
\end{equation}%
where $d_0$ corresponds to the coefficient $\chi^{(2)}$and the respective duty cycle is $1/2$ \cite{Aviv2018}. We retain, as usual, a truncated form of expansion (\ref{QPM}),
with only the fundamental harmonics corresponding to $m=1$ and $m=-1$ kept
in Eqs. (\ref{A1}) and (\ref{A2}), respectively, as they play the dominant
role in QPM.

Natural rescaling (cf. Ref. \cite{Phillips,ZFY}),
\begin{gather}
{{I}_{0}}=\left( \frac{{{n}_{1}}}{{{\omega }_{1}}}+\frac{{{n}_{2}}}{{{\omega
}_{2}}}\right) {{\left\vert {{A}_{0}}\right\vert }^{2}},\quad z_{d}^{-1}=%
\frac{2d_{0}{{\omega }_{1}}}{\pi c{{n}_{1}}}\sqrt{{{\frac{{{\omega }_{2}}}{{{%
n}_{2}}}{{I}_{0}}}}}{,}  \label{zd_A0} \\
\psi _{p}={{A}_{p}}\sqrt{\frac{{{n}_{p}}}{{{\omega }_{p}I}_{0}}}{\exp }\left[
i(\Delta k-2\pi /\Lambda )Z\right] ,\quad p=1,2,  \label{psi} \\
z=Z/z_{d},\quad x=X\sqrt{{{k}_{1}}/z_{d}},\quad y=Y\sqrt{{{k}_{1}}/z_{d}},
\label{normalization} \\
\Omega =z_{d}(\Delta k-2\pi /\Lambda ),\quad  \label{Delta}
\end{gather}%
where $A_{0}$ is a characteristic amplitude of the electromagnetic field,
and $\Omega $ is the effective detuning, casts Eqs. (\ref{A1}) and (\ref{A2}%
), with the truncated form of expansion (\ref{QPM}), in the normalized form,
with neglected difference between $n_{1}$ and $n_{2}$ (which is relevant for
available materials):
\begin{eqnarray}
&&i\partial _{z}\psi _{1}=-\frac{1}{2}\nabla _{x,y}^{2}\psi _{1}-\Omega \psi
_{1}-2 \sigma(x,y) \psi _{1}^{\ast }\psi _{2},  \label{psi1} \\
&&i\partial _{z}\psi _{2}=-\frac{1}{4}\nabla _{x,y}^{2}\psi _{2}-\Omega \psi
_{2}-\sigma(x,y)\psi _{1}^{2},  \label{psi2}
\end{eqnarray}%
and $\nabla _{x,y}^{2}=\partial _{x}^{2}+\partial _{y}^{2}$. Equations (\ref%
{psi1}) and (\ref{psi2}) conserve two dynamical invariants, \textit{viz}.,
the total Hamiltonian and power (alias the Manley-Rowe invariant \cite{Gil}%
),
\begin{eqnarray}
&&H=\iint {(\mathcal{H}_{P}+\mathcal{H}_{\Omega }+\mathcal{H}_{\chi ^{(2)}})}%
dxdy  \label{H} \\
&&P=\iint (|\psi _{1}|^{2}+2|\psi _{2}|^{2})dxdy\equiv P_{1}+P_{2}.
\label{Norm}
\end{eqnarray}%
where $\mathcal{H}_{P}=\frac{1}{2}|\nabla {{\psi }_{1}}|^{2}+\frac{1}{4}%
|\nabla {{\psi }_{2}}|^{2}$, $\mathcal{H}_{\Omega }=-\Omega (|\psi
_{1}|^{2}+|{\psi _{2}|}^{2})$, and $\mathcal{H}_{\chi ^{(2)}}=-\sigma(x,y)%
\left[\left( {\psi _{1}^{\ast }}\right) ^{2}{\psi _{2}}+\mathrm{c.c}\right]$%
. Power sharing between the FF and SH components is characterized by ratio $%
\gamma =P_{2}/P_{1}$. Control parameters of the resulting system are $P$, $%
\Omega $, and $D$.

\begin{figure}[h]
{\includegraphics[height=0.36\columnwidth]{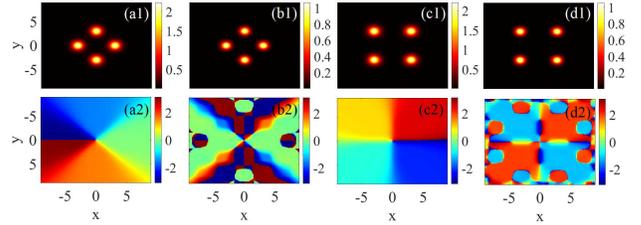}}
\caption{Generic examples of stable rhombus-shaped and square-shaped vortex
solitons composed of four peaks subject to condition $\protect\varphi _{2}=2%
\protect\varphi _{1}$ or $\protect\varphi _{2}=2\protect\varphi _{1}-\protect%
\pi $, respectively. These modes are centered at a white square cell, as
shown in Fig. \protect\ref{crystal}(a). Panels (a1,b1) and (c1,d1) display
the distribution of the local intensity of the FF and SH components of the
rhombus- and square-shaped vortex solitons, respectively. Panels (a2-d2)
present the phase distribution of the solitons corresponding to (a1-d1),
respectively. The parameters here are $(P,D,\Omega )=(40,3,0)$.}
\label{rhombus}
\end{figure}

Bright-vortex soliton solutions to Eqs. (\ref{psi1}) and (\ref{psi2}) are
looked for as
\begin{equation}
\psi _{p}(x,y,z)=\phi _{p}(x,y)\exp (ip\beta z),p=1,2,  \label{exp}
\end{equation}%
where $\phi _{1,2}$ represent the stationary shape of the FF and SH
components, with propagation constants $\beta $ and $2\beta $, respectively.
The $\chi ^{(2)}$ terms in Eqs. (\ref{psi1}) and (\ref{psi2}) impose the
matching condition on phases $\varphi _{1,2}\left( x,y\right) =\mathrm{Arg}%
\left\{ \phi _{1,2}\left( x,y\right) \right\} $ of the stationary solution:
\begin{equation}
\varphi _{2}\left( x,y\right) =2\varphi _{1}\left( x,y\right) -\varphi
_{d}\left( x,y\right), \label{matching}
\end{equation}%
where $\varphi_{d}(x,y)$ is decided by $\sigma(x,y)$. $\sigma(x,y)=1$ and $-1$ correspond to $\varphi_{d}(x,y)=0$ and $\pi$, respectively.
It implies that there may exist two different types of the vortex solitons.
Each one is built, essentially, of four peaks (local intensity maxima),
which form rhombic or square patterns, as shown in Fig. \ref{rhombus}. In
these patterns, the vorticity is represented by the phase circulation along
a closed path linking the peaks, which is the generic method for creating
the multi-peak vortices supported by underlying lattice structures \cite%
{BBB,Musslimani} (the vorticity may be defined in this way even if the
spatially modulated system does not conserve the angular momentum). As a
result, the rhombic and square-shaped patterns obey the matching condition (%
\ref{matching}) with $\varphi _{2}\left( x,y\right) =2\varphi _{1}\left(
x,y\right) $ or $\varphi _{2}\left( x,y\right) =2\varphi _{1}\left(
x,y\right) -\pi $. The matching condition does not hold in the empty square
cell at the center of the rhombic pattern, hence this pattern may be
classified as \textit{intersite-centered} one \cite{Kevr}, while the cell at
the square's center may satisfy the matching, but is left empty, thus
representing an \textit{onsite-centered} self-trapped state. Accordingly,
the rhombuses and squares are built, respectively, as densely and loosely
packed structures.

Thus, the phase patterns of the two-component vortex solitons demonstrate,
in Figs. \ref{rhombus}(a2,c2), that the FF field carries the vorticity with
winding number $1$. As concerns the SH component, its phase is plotted in
panels (b2,d2) on scale $-2\pi <\varphi _{2}<+2\pi $, subtracting $\pm 2\pi $
in the region where Eq. (\ref{matching}) yields $\left\vert \varphi
_{2}\right\vert >2\pi $. For this reason, the phase patterns in these panels
seem as corresponding to a quadrupole mode. Numerical results also
demonstrate that, unlike previously studied multi-peak vortex solitons
supported by linear square-shaped potentials \cite{BBB,Musslimani,book}, the
present patterns are very sharp ones, in the sense that they do not feature
side peaks in lattice cells which do not belong to the main rhombus or
square. This feature may be useful for potential applications.

\begin{figure}[tbh]
{\includegraphics[height=0.36\columnwidth]{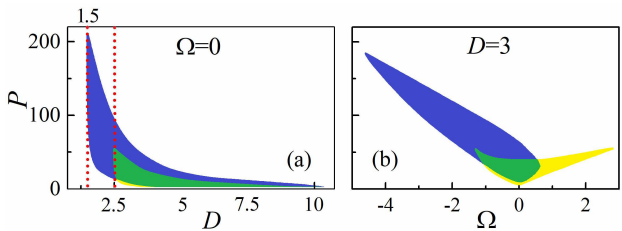}}
\caption{The rhombic and square-shaped vortex solitons are stable,
separately, in blue and yellow areas in the $(P,D)$ parameter plane with $%
\Omega =0$ (a), and in the $(P,\Omega )$ plane with $D=3$ (b). They coexist
as stable states in green areas in both planes. Vertical red dotted lines in
(a) designate existence boundaries for the two soliton species.}
\label{stablearea}
\end{figure}

\begin{figure}[h]
{\includegraphics[height=0.45\columnwidth]{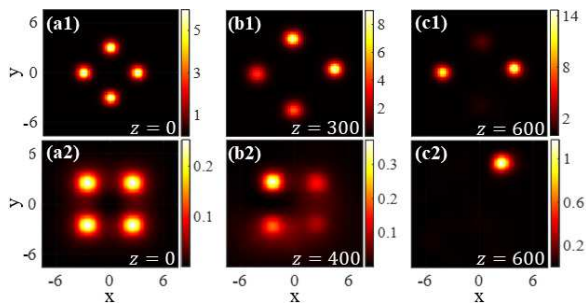}}
\caption{Typical examples of the evolution of unstable solitons, which are
selected outside of the stability area in Fig. \protect\ref{stablearea}.
(a): Intensity shapes of unstable rhombic and square vortex solitons with $(P,D,\Omega
)=(70,3,0)$ and $(5,5,0)$, plotted at $z=0$ (a1,a2), $300$ and $400$ (b1,b2), and $600$ (c1,c2), respectively.}
\label{unstable}
\end{figure}

Stability of the vortex solitons was verified by direct simulations of the
perturbed evolution in the framework of Eqs. (\ref{psi1}) and (\ref{psi2}),
up to $z=1000$. The resulting stability areas in the $(P,D)$ and $(P,\Omega
) $ planes are displayed in Fig. \ref{stablearea}. Comparison of the results
for the two types of the vortex solitons shows that the rhombic species has
a wider stability area than its square-shaped counterpart. The difference is
explained by the fact that the above-mentioned tight structure of the
rhombuses helps to stabilize them stronger than it is possible with the help
of the loose binding in squares. Indeed, the solitons of the former type
remain stable up to value $P_{\max }\approx 200$ of the total power, while
the square-shaped family has $P_{\max }\approx 60$. Intervals of the
detuning parameter for the stable rhombuses and squares are, respectively, $%
-4.6<\Omega <0.6$ and $-1.3<\Omega <2.8$. Furthermore, there is a minimum
size of spatial period $D$ of the checkerboard lattice necessary for
supporting rhombuses and squares, \textit{viz}., $D_{\min }^{\text{\textrm{%
(rhomb)}}}=1.5$ and $D_{\min }^{\text{\textrm{(square)}}}=2.5$, as shown by
the vertical dotted red lines in Fig. \ref{stablearea}(a). The compound
vortex solitons do not exist at $D<D_{\min }$, as the rapid alternation of
the sign in front of the $\chi ^{(2)}$ terms in Eqs. (\ref{psi1}) and (\ref%
{psi2}) leads to effective cancellation of the quadratic nonlinearity. At $%
D>D_{\min }$, those rhombic and square-shaped solitons which are unstable
spontaneously decay into sets of one or two peaks, as shown in Fig. \ref%
{unstable}.

\begin{figure}[h]
{\includegraphics[height=0.55\columnwidth]{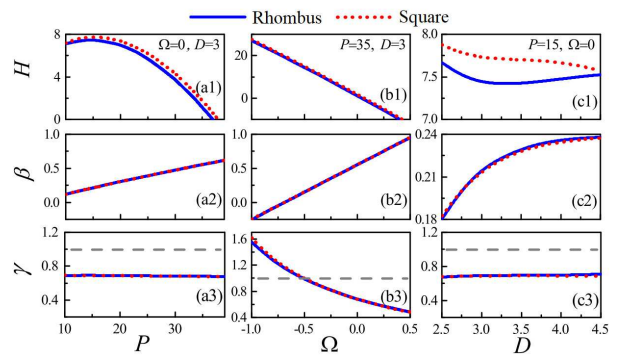}}
\caption{Hamiltonian $H$, propagation constant $\protect\beta $, and SH/FF
power ratio $\protect\gamma $ for the vortex-soliton families vs. total
power $P$, mismatch $\Omega $, and spatial period $D$ of the underlying
lattice. The horizontal dashed lines in the bottom panels designate the
level of $\protect\gamma =1$. Solid and dotted curves represent the rhombic-
and square-shaped solitons, respectively. Parameters are fixed as $(\Omega
,D)=(0,3)$, $(P,D)=(35,3)$, and $(P,\Omega )=(15,0)$ in the left, middle,
and right columns of panels, respectively. At these parameters, the curves
in all the panels belong to the stability areas, see Fig. \protect\ref%
{stablearea}.}
\label{char}
\end{figure}

Families of the vortex solitons considered here are quantified by
dependences of their characteristics, $H$, $\beta $, and $\gamma $, on $P$, $%
\Omega $ and $D$, as shown in Fig. \ref{char}. In particular, the top row
shows that the rhombic solitons realize the system's ground state, as their
value of the Hamiltonian is lower than that of the square-shaped ones. This
property explains why the rhombuses have larger stability areas in Fig. \ref%
{stablearea}. Further, Fig. \ref{char}(a2) shows that the soliton families
satisfy the Vakhitov-Kolokolov criterion, $d\beta /dP>0$, which is a
well-known necessary stability condition for solitons in self-focusing media
\cite{VK}. The bottom row in Fig. \ref{char} shows that the power-ratio
factor $\gamma $ depends solely on mismatch $\Omega $, but not on $P$ and $D$%
. Furthermore, Fig. \ref{char}(b3) shows that $\Omega =-0.5$ is a boundary
between the two-component vortex solitons dominated by the SH and FF
components (i.e., $\gamma >1$ and $\gamma <1$) at $\Omega <-0.5$ and $\Omega
>-0.5$, respectively.

\begin{figure}[h]
{\includegraphics[height=0.4\columnwidth]{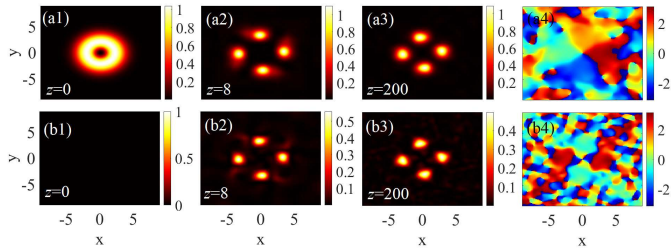}}
\caption{The simulated propagation of the input launched in the FF component
at $z=0$ in the form a Laguerre-Gaussian vortex beam with $P=40$ and winding
number $1$. (a1-a3) and (b1-b3): The intensity pattern of the FF and SH
components are displayed at $z=0$ (a1,b1), $8$ (a2,b2) and $200$ (a3,b3),
respectively. (a4,b4): The FF phase pattern of FF and SH components at $%
z=200 $, respectively. Here, $D=3$ and $\Omega =0$ are fixed.}
\label{Evolution}
\end{figure}

To verify the experimental feasibility of the present setup, the creation of
stable two-component vortex solitons from the natural input in the form of
the standard Laguerre-Gaussian vortex beam with winding number 1, initially
launched into the checkerboard photonic crystal with length $z=200$ in the
FF component, was simulated too. Figure \ref{Evolution} shows a typical
example, which demonstrates the formation of the well-defined soliton after
passing $z\approx 8$. Naturally, the distance necessary for the creation of
the robust output decreases with the increase of the input power. A
schematic diagram of the experimental setup in which the above results may
be realized is displayed above in Fig. \ref{crystal}(d). We also attempted to construct vortex-soliton solutions with winding number $2$ in the FF component. However, the numerical analysis has not produced any such stable solutions \cite{Additionalsimulaiton}.

To estimate experimentally relevant characteristics for the predicted
compound solitons, we adopt parameters for lithium niobate ($d_{0}=27$ pm/V
\cite{NLO}, $n_{1}\approx n_{2}\approx 2.2$). Experimentally relevant values
of the FF and SH\ wavelengths, and amplitude of the electric field $A_{0}$
in Eq. (\ref{zd_A0}) are chosen as $1064$ nm and $532$ nm, and $200$ kV/cm
\cite{1Buryak2002}, respectively, which yields the propagation-distance
scale $z_{d}=0.0625$ cm [cf. Eq. (\ref{zd_A0})]. Table I summarizes
relations between the variables in the scaled and physical units.

\begin{table}[h]
\caption{The relation between the coordinates, intensities, and power
measured in scaled and physical units.}
\label{tableI}%
\begin{ruledtabular}
\begin{tabular}{lcr}
\specialrule{0em}{1pt}{1pt}
$z=1$ \& $\Omega=1$ &$\quad$ &  0.0625 cm \& 16 cm$^{-1}$\\
\specialrule{0em}{1pt}{1pt}
$x=1$  \& $y=1$ &$\quad$ & 7 $\mathrm{\mu}$m \\
\specialrule{0em}{1pt}{1pt}
$|\psi_{1}|^2=1$ \& $|\psi_{2}|^2=1$  &$\quad$ & 80 MW/cm$^{2}$ \& 160 MW/cm$^{2}$ \\
\specialrule{0em}{1pt}{1pt}
$P=1$ & $\quad$ & 40 W\\
\specialrule{0em}{1pt}{1pt}
\end{tabular}
\end{ruledtabular}
\end{table}

With the values from Table I, the total power for the solitons displayed in
Fig. \ref{rhombus} is $1600$ W, peak intensities of both components being $%
\approx 0.17$ GW/cm$^{2}$. The\ physical propagation length corresponding to
$z=1000$ is $62.5$ cm, which is tantamount to hundreds of the Rayleigh
(diffraction) lengths, corroborating the complete stability of the solitons.
We also note that, because the $\chi ^{(3)}$ coefficient for the same
material is $36.6\times 10^{-22}$ m$^{2}$/V$^{2}$ \cite{Kulagin}, the Kerr
nonlinearity is indeed negligible in the present setting for the current
soliton intensity. Further, the spatial period of the checkerboard in the
range of $1.5\leq D\leq 10$ corresponds to the width of the square cell
ranging from $10$ $\mathrm{\mu }$m to $70$ $\mathrm{\mu }$m in physical
units. In particular, zero mismatch, $\Omega =0$, is attained at $\Lambda
\approx 6.65$ $\mathrm{\mu }$m (the refractive index difference for the FF
and SH components with the type-I polarization is $\Delta n\approx 0.08$
\cite{Jundt}, producing, as mentioned above, a negligible effect in the
present setup). These length scales are realistic for the implementation
using the available QPM-fabrication technique. In physical units, Fig. \ref%
{Evolution} demonstrates that the input in the form of the Laguerre-Gaussian
vortex beam with power $1600$ W and FWHM width $20$ $\mathrm{\mu }$m the
soliton is formed at $Z\approx 0.5$ cm (which corresponds to the scaled
distance $z\approx 8$). The respective experimental setup is shown above in
Fig. \ref{crystal}(d). The appropriate Laguerre-Gaussian laser beam at
wavelength $\lambda =1064$ nm can be generated by a high-power laser source,
e.g., a Nd:YAG system generating pulses with energy and temporal width 10 $%
\mathrm{\mu }$J and 6 ns, at $40$ Hz repetition rate. The half-wavelength
plane (HWP), adjusted to the type-I polarization type, and the vortex-phase
plates (VPP) morph the input into a vortex beam with orbital angular
momentum $l=1$, which is focused by the lens onto the photonic crystal.
Then, it evolves as shown in Fig. \ref{Evolution}. According to the above
estimates, the optical multi-peak vortex soliton can be generated and
detected by the CCD in the output beam, provided that the crystal's length
exceed $0.5$ cm.

Losses of the optical medium may be considered as well. According to Refs.
\cite{Kashiwazaki}, we assume that the loss rate for both FF and SH waves in
lithium niobate is $\approx 1.56\%$ per centimeter. Adding the linear losses
to Eqs. (\ref{psi1}) and (\ref{psi2}), we have checked that the system still
predicts the formation of robust vortex solitons. For example, the rhombic
and square-shaped solitons observed at $Z=20$ cm keep more than $70\%$ of
the initial power.


In conclusion, we have produced two families of compound vortex solitons,
rhombic and square-shaped ones (which are intersite- and onsite-centered
self-trapped states, respectively), in the $\chi ^{(2)}$ medium equipped
with the QPM checkerboard lattice. Unlike completely unstable vortex
solitons in spatially uniform $\chi ^{(2)}$ systems, the presently
considered families feature broad stability areas in the system's parameter
space. The tightly-bound rhombic vortex solitons realize the system's ground
state, featuring the lowest value of the Hamiltonian and a larger stability
area than loosely bound squares. Physical characteristics of the predicted
vortex-soliton families belong to the range accessible to currently
available experimental techniques.

The analysis can be extended in other directions. In particular, a natural
objective is to consider a chirped QPM, corresponding to $Z/\Lambda $
replaced by $\int dZ/\Lambda (Z)$ in Eq. (\ref{QPM}), with variable spatial
period $\Lambda (Z)$, which makes detuning $\Omega $ a function of $Z$ in
Eq. (\ref{Delta}) and suggests possibilities to manipulate the solitons, cf.
Ref. \cite{Aviv2022}. On the other hand, if the Fourier components with $%
|m|>1$ are taken into account, an effective cubic nonlinearity with self-
and cross-phase modulation can be induced through the cascading mechanism.
It may also be relevant to include the intrinsic $\chi ^{(3)}$ nonlinearity
into the consideration. Effects of the cubic terms on the stability of
vortex solitons is an interesting issue \cite{OBang,10Mihalache}. A
challenging problem is mobility of 2D vortex solitons against the lattice
backdrop. Previously, mobility was demonstrated only for fundamental 2D
lattice solitons in the $\chi ^{(2)}$ system \cite{Susanto}.

\begin{acknowledgments}
This work was supported by the NNSFC (China) through Grants No. 12274077,
11874112, 11905032, 62005044, by the Guangdong Basic and Applied Basic
Research Foundation through grant No. 2021A1515111015, the Key Research
Projects of General Colleges in Guangdong Province through grant No.
2019KZDXM001, the Research Fund of Guangdong-Hong Kong-Macao Joint
Laboratory for Intelligent Micro-Nano Optoelectronic Technology through
grant No.2020B1212030010 and the Graduate Innovative Talents Training
Program of the Foshan University. The work of B.A.M. is supported, in part,
by the Israel Science Foundation through grant No. 1695/22.
\end{acknowledgments}


\begin{thebibliography}{99}
\bibitem{Additionalsimulaiton} Additional numerical results demonstrate that
modes with a quadrupole structure (instead of the vortex) in the FF
component exist and some of them are stable.
\end{thebibliography}
\end{document}